\documentclass[conference]{IEEEtran}
\IEEEoverridecommandlockouts
\usepackage{comment}
\usepackage{ulem}
\usepackage{cite}
\usepackage{framed}
\usepackage{color}
\usepackage{slashbox}
\usepackage{listings}
\usepackage{latexsym}
\usepackage{graphicx}
\usepackage{url}

\usepackage{algorithm}
\usepackage{algpseudocode}

\usepackage[fleqn]{amsmath}
\usepackage{amssymb}
\usepackage{amsfonts}

\def\BibTeX{{\rm B\kern-.05em{\sc i\kern-.025em b}\kern-.08em
    T\kern-.1667em\lower.7ex\hbox{E}\kern-.125emX}}
    
\begin{document}

\title{Formal Verification of Decision-Tree Ensemble Model and Detection of its Violating-input-value Ranges
}


\author{
\IEEEauthorblockN{Naoto Sato\IEEEauthorrefmark{1}, Hironobu Kuruma\IEEEauthorrefmark{1}, Yuichiroh Nakagawa\IEEEauthorrefmark{1}, Hideto Ogawa\IEEEauthorrefmark{1}}
\IEEEauthorblockA{\IEEEauthorrefmark{1}Research \& Development Group, Hitachi, Ltd.}
}

\maketitle

\begin{abstract}
As one type of machine-learning model, a ``decision-tree ensemble model" (DTEM) is represented by a set of decision trees. A DTEM is mainly known to be valid for structured data; however, like other machine-learning models, it is difficult to train so that it returns the correct output value for any input value. Accordingly, when a DTEM is used in regard to a system that requires reliability, it is important to comprehensively detect input values that lead to malfunctions of a system (failures) during development and take appropriate measures. One conceivable solution is to install an input filter that controls the input to the DTEM, and to use separate software to process input values that may lead to failures. To develop the input filter, it is necessary to specify the filtering condition of the input value that leads to the malfunction of the system. Given that necessity, in this paper, we propose a method for formally verifying a DTEM and, according to the result of the verification, if an input value leading to a failure is found, extracting the range in which such an input value exists. The proposed method can comprehensively extract the range in which the input value leading to the failure exists; therefore, by creating an input filter based on that range, it is possible to prevent the failure occurring in the system. In this paper, the algorithm of the proposed method is described, and the results of a case study using a dataset of house prices are presented. On the basis of those results, the feasibility of the proposed method is demonstrated, and its scalability is evaluated. 
\end{abstract}


\section{Introduction}\label{sec_intro}
Recently, software developed by machine learning has been used in various systems. Deep learning using deep neural networks (DNNs) is widely used for predicting and classifying image data, audio data \cite{dl_image}\cite{dl_speech}, and so on. For structured data, ensemble learning methods using decision trees, such as random forests \cite{rf} and gradient-boosting decision trees \cite{gbdt}, are also effective \cite{gbdt_ex1}\cite{gbdt_ex2}\cite{gbdt_ex3}\cite{xgboost}\cite{rf_ex1}\cite{rf_ex2}\cite{rf_ex3}\cite{rf_ex4}. 

A decision-tree ensemble model (DTEM) is represented as a set of decision trees. The prediction result output from a DTEM is calculated as the sum or average value of the scores associated with the leaves of the decision trees. The DTEM is expected to be generalized, namely, to return the appropriate output value even if it is given an input value not included in the training data.

\begin{figure}[htb]
\begin{center}
\scalebox{0.55}{\includegraphics[bb=0 0 405 125]{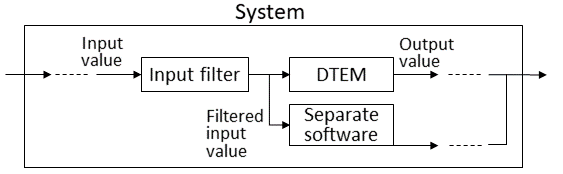}}
\end{center}
\caption{Input filter}
\label{fig01}
\end{figure}

However, in general, it is difficult to train a DTEM to return the appropriate output value for every input value, that is, a DTEM returns an inappropriate output value with a certain probability. Therefore, in particular, when a DTEM is used in a mission-critical system, whose behavior significantly affects business and society, it is important to comprehensively detect input values that lead to system failures during development, and take appropriate measures. As countermeasures, retraining or additional training of the DTEM are possible means; however, for the reason mentioned above, it is difficult to completely eliminate the possibility of failures occurring. Accordingly, as a practical measure, it is possible to create an input filter to control the input value to the DTEM and to use separate software to process the filtered input values that leads to failures (Fig. \ref{fig01}). The implementation of the separate software is arbitrary. For example, the input value might be rejected to be processed. 

A policing function \cite{policing} is also useful to prevent failures. It is separated from the DTEM, and checks the output value of the DTEM againtst a certain property at runtime. If the output value does not meet the property, that is, the output value leads to failures, the policing function controls the output value. However, when comparing the input filter and the policing function, the input filter is more efficient because the input filter detects and controls failures before the DTEM runs. 

To create the input filter, it is necessary to specify which input values should be filtered as the filtering condition. In this paper, we therefore propose a method to formally verify whether a DTEM meets certain property and, if that properties is not satisfied, extracts the range (part of the input-value space) in which all input values violating the property are included. By setting the range extracted by the proposed method as the filtering condition of the input filter, it is possible to prevent the failure of the system due to the DTEM. 
The feasibility of the proposed method is evaluated by showing a case study in which the proposed method was applied experimentally. Moreover, the scalability of the proposed method is evaluated by changing the dimension of the input values of the DTEM, the number of decision trees constituting the DTEM, and the maximum value of the depth of those decision trees. Hereafter, input values that violate the property are referred to as ``violating input values." And the range in which the violating input value exists is called the ``violation range." It should be noted that not all input values included in the violation range will be violating input values.

As for the proposed method, the decision trees that compose the DTEM are encoded to a formula, and the formula is verified by using a satisfiability modulo theories (SMT) solver. Although this approach has mainly been applied to DNNs \cite{stanford_lp1}\cite{stanford_lp2}\cite{stanford_lp3}\cite{planet}\cite{oxford_smt}\cite{unified}, to the authors' knowledge, no case in which this approach is applied to a DTEM has been reported. Given that situation, the first contribution of this paper is to demonstrate that our approach, namely, logically encoding a machine-learning model and verifying the model by solving the resulting a formula with an SMT solver, is also applicable to a DTEM.

When the DTEM violates a property, as a result of the verification, an example of a violating input value (and its corresponding output value)---called a ``counterexample"---is obtained. As a naive way to get the filtering condition, all violating input values are detected by repeating the verification. If an input value matches any violating input values, it is filtered. However, a large number of similar violating input values may exist around a certain violating input value. In particular, if the input value is represented by a multidimensional vector whose elements are numeric variables (not categorical variables), input values obtained by slightly changing the values of some elements of the violating input value are also likely to violate the property. In this case, since the verification is repeated as many times as the number of the violating input values, detecting all the violating input values is not practical in terms of calculation time.

Targeting a DTEM whose input values are multi-dimensional vectors whose elements are numeric variables, the proposed method extracts the violation range by searching around the origin at which a violating input value was first detected and gradually expanding the search range until the violating input value is not detected. With this method, the violating input values around the origin can be kept within the range. However, input values that do not violate the property are also included in the range. Even though they do not violate the property, they are also filtered as well as the violating input values.
Thus, it is desireble to prevent as many input values that do not violate the property as possible from being included in the violation range.
Therefore, as for the proposed method, the extracted violation range is divided into a number of smaller ranges. Then, for each divided range, whether a violating input value exists within the range is checked. As a result, it is possible to narrow down the original violation range. 
As for the second contribution of this paper, targeting a DTEM taking multi-dimensional vectors of numerical variables as input values, we propose a method to extract the violation range and narrow it down. Moreover, the feasibility and scalability of the proposed method are evaluated through a case study. 

The rest of this paper is organized as follows. In Section \ref{sec_prelim}, a decision tree and a DTEM are formally defined. In Section \ref{sec_overview}, the proposed method is overviewed. In Section \ref{sec_fv}, among the procedures that compose the proposed method, the procedure for verifying the DTEM is explained. In Section \ref{sec_extract}, the procedure for extracting the violation range on the basis of the verification result is explained. In Section \ref{sec_divide}, the procedure for narrowing down by dividing the extracted violation range is explained. In Section \ref{sec_casestudy}, the feasibility and scalability of the proposed method are evaluated through a case study using a data set of house prices. In Section \ref{sec_discuss}, the usefulness and applicability of the proposed method are discussed. In Section \ref{sec_relwork}, related work is described, and in Section \ref{sec_conclusion}, the conclusions drawn from this study are presented.

\section{Preliminaries}\label{sec_prelim}
For any DTEM $M$, variables representing input and output values of $M$ are given as $x$ and $y$, respectively. It is assumed that $x$ is represented by a vector of length $s \geq 2 $ such that $[x[0], ..., x[k], ..., x[s-1]]$. As for $M$ targeted in this paper, since multi-dimensional vectors with numeric variables as elements are assumed as input values, $x[0], ..., x[k], ..., x[s-1]$ represents each numeric variable. And the domain of $M$ is represented as $X$. First, in Section \ref{subsec_tree}, the decision trees that compose $M$ are formally defined. Next, in Section \ref{subsec_dtem}, $M$ is formally defined.

\subsection{Decision Tree}\label{subsec_tree}
\begin{figure}[htb]
\begin{center}
\scalebox{0.45}{\includegraphics[bb=0 0 400 283]{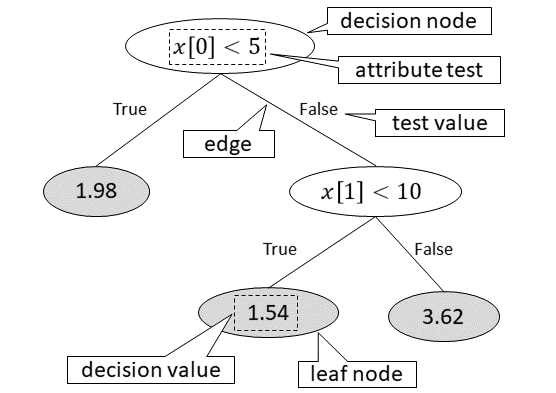}}
\end{center}
\caption{Structure of decision tree}
\label{fig09}
\end{figure}

A decision tree represents a procedure for determining the class to which a given input value belongs \cite{decisiontree1}\cite{decisiontree2}. A decision tree consists of decision nodes, edges, and leaf nodes. The expression of input value $x$ is associated with the decision node (Fig. \ref{fig09}). In this paper, the expression of the decision node is called an ``attribute test." Each decision node has multiple child nodes, and each child node is connected by an edge. Each edge is labeled with the evaluation result of the attribute test. When an input value is given, the edge with the same label as the result of evaluating the attribute test is selected. Hereafter, the evaluation result of the attribute test is called the ``test value." 
If the child node connected to the selected edge is a decision node, the same procedure is performed for that decision node. If the child node is a leaf node, the value associated with that leaf node becomes the name of the class to which the input value belongs. As for the decision trees making up the DTEM, numeric values are used as class names, and those numeric values are used to calculate the output value of the DTEM. Hereafter, this value is called ``the decision value."

Arbitrary decision tree $t$ constituting $M$ can be expressed in the form $(N_{d}, N_{l}, n_{1}, E, attr, tv, dv)$, where $N_{d}$ represents a set of decision nodes, $N_{l}$ represents a set of leaf nodes, and $n_{1}$ represents a root node included in $N_{d}$. $E$ is a set of edges, each of which edge is represented by a pair consisting of a connection-source node and a connection-destination node. That is, it can be defined as $E \subseteq N_{d} \times (N_{l} \cup (N_{d} \setminus \{n_{1}\}))$. Here, $attr: N_{d} \longrightarrow A$ is a function that associates an attribute test with a decision node. $A$ represents a set of arbitrary expressions for input value $x$. Similarly, $tv: E \longrightarrow V$ is a function that associates a test value with an edge, where $V$ represents a set of test values. In the example in Fig. \ref{fig09}, $V=\{True, False \}$. Here, $dv: N_{l} \longrightarrow \mathbb{R}$ is a function that associates a decision value with a leaf node. The decision tree is acyclic. This is defined, by using transitive closure $E^{+}$ of $E$, as $ \forall n \in N_{d} \cup N_{l} \cdot \langle n, n \rangle \not \in E^{+}$. 

It is assumed that the decision tree that constitutes the DTEM, which is the subject of this paper, has more than one decision nodes, and branches from a decision node to at least two child nodes. Therefore, if the number of nodes included in $N_{d}$ is expressed as $card(N_{d})$, $card(N_{d}) \geq 1$ holds. In a similar manner, $card(N_{l}) \geq 2$ is established. As for the number of edges, $card(E)$, $card(E) \geq 2$ holds.

\subsection{Decision Tree Ensemble Model(DTEM)}\label{subsec_dtem}
Arbitrary $M$ can be expressed as $(T, ensem)$. $T$ represents a set of decision trees constituting $M$. If the number of decision trees included in $T$ is taken as $card(T)$, the decision trees included in $T$ can be expressed as $t_{1}, ..., t_{i}, ..., t_{card(T)}$. The decision values of trees $t_{1}, ..., t_{i}, ..., t_{card(T)}$ are represented as $y_{1}, ..., y_{i}, ..., y_{card(T)}$, respectively. 

Moreover, $ensem$ represents a function that takes $y_{1}, ..., y_{card(T)}$ as arguments and returns $y$. The specification of $ensem$ depends on the implementation of DTEM. For example, in the case of random forests, the average of the decision values is $y$. Alternatively, in the case of a gradient-boosting decision tree, the sum of the decision values is $y$. In this paper, the calculations depending on the implementation of the DTEM are generalized with $ensem$.

\section{Outline of Proposed Method}\label{sec_overview}
The proposed method is outlined in Algorithm \ref{al1}. As for the proposed method, $M$ is taken as the DTEM to be verified. A property of $M$ is denoted by $ \varphi $, and $X$ represents a domain of $M$. Parameters $r_{a}$, $r_{b}$, and $r_{c}$ are described later. Algorithm \ref{al1} outputs the value of variable $vranges$, which is a set of violation ranges for $ \varphi $. 
\begin{algorithm}[htb]
\caption{Violation Ranges Detection}
\label{al1}
\begin{algorithmic}[1]
\Require $M$, $ \varphi $, $X$, $r_{a}$, $r_{b}$, $r_{c}$
\Ensure $vranges$
\newline

\State $vranges$ $\leftarrow$ $\emptyset$
\State $\rho$ $\leftarrow $ $input\_is\_within$($M$, $X$) \label{dom}
\While{True}
  \State $ce$ $\leftarrow$ $formal\_verification$($M$, $ \varphi $, $\rho$) \label{veri}
  \If{$ce$ $\neq$ None}
    \State $vio$, $novios$ $\leftarrow$ $range\_extraction$($ce$, $M$, $ \varphi $, $\rho$, $r_{a}$, $X$) \label{search}
    \State $core \leftarrow vio$ \label{copy_range}
    \While{$continue\_division$($core$, $X$, $r_{b}$)\label{necess} \\
      \hspace{4.5cm} $\land$ $novios$ $\neq$ $\emptyset$}
      \State $iv$ $\leftarrow$ $novios$.pop()
      \State $core$, $surrds$ $\leftarrow$ $range\_division$($core$, $iv$, $\rho$, $r_{c}$, $M$, $ \varphi $) \label{split}
      \State $vranges$ $\leftarrow$ $vranges$ $\cup$ $surrds$ \label{add_sur}
    \EndWhile
    \State $vranges$ $\leftarrow$ $vranges$ $\cup$ $core$ \label{fin_split}
    \State $\rho$ $\leftarrow$ $\rho$ $\land$ $input\_is\_outside$($M$, $vio$) \label{other_range}
  \Else
    \State break
  \EndIf
\EndWhile
\State \Return $vranges$
\end{algorithmic}
\end{algorithm}
Function $input\_is\_within$ on line \ref{dom} returns formula $\rho$, meaning that ``Input value $x$ of $M$ is included in domain $X$." Procedure $formal\_verification$ on line \ref{veri} verifies whether an input value that violates $ \varphi $ and satisfies constraint $ \rho $. If such an input value is detected, that value is assigned to variable $ce$. If no such input value is detected, ``None" is assigned to $ce$. Procedure $formal\_verification$ is described in detail in Section \ref{sec_fv}. Procedure $range\_extraction$ on line \ref{search} extracts the violation range for $ \varphi $ from around $ce$. The extracted range is assigned to variable $vio$. Moreover, a range in which no input value violates $ \varphi $ is extracted by the process of extracting $vio$; therefore, a set of such ranges is output as variable $novios$. Hereafter, a range in which no violating input value exists is referred to as a ``no-violation range." Procedure $range\_extraction$ is shown in detail in Section \ref{sec_extract}.

The extracted violation range, $vio$, is divided as follows. First, on line \ref{copy_range}, $vio$ is copied to variable $core$. On line \ref{necess}, $continue\_division$ is a function that returns a decision of whether to execute procedure $range\_division$ on line \ref{split}. Parameter $r_{b}$ is used as a reference value for the decision. Although the details of $continue\_division$ are not specified here, for example, it can be implementated so as to return ``True" if the volume of core is equal to or more than $r_{b}$ percent of the volume of $X$. On line \ref{split}, procedure $range\_division$ divides $core$ into multiple pieces. This division is performed on the basis of a no-violation range, which is an element of $novios$. Of the ranges obtained by division, the inner range including $ce$ (which is the starting point of range extension) is taken as a new $core$, and the set of other outer ranges is taken as $surrds$. The $range\_division$ is explained in detail in Section \ref{sec_divide}.

If $novios$ contains a number of no-violation ranges, it is used for the division from the range pushed to $novios$ last, which is the outermost no-violation range in terms of violating input value $ce$.
Therefore, the ranges stored in $surrds$ do not include no-violation ranges because they are outside the outermost no-violation range. Accordingly, the elements of $surrds$ are added to $vranges$ without dividing them further (line \ref{add_sur}). On the contrary, the new $core$ might include a no-violation range. Therefore, on line \ref{necess}, it is evaluated by $continue\_division$ whether $core$ will be divided further. If $core$ is not divided any more, it is added to $vranges$.

Then, on line \ref{other_range}, function $input\_is\_outside$ is used to create a constraint representing ``Input value x of $M$ is out of the range indicated by $vio$.", and that constraint is conjunctively appended to $ \rho $. After that, by re-executing formal\_verification on line \ref{veri}, it is verified whether an input value violates $ \varphi $ outside $vio$. If an input value that violates $ \varphi $ is detected as a result of the verification, the violation range is extracted and divided by the same procedure. 
$vranges$ extracted by Algorithm \ref{al1} represents the range in which an input value violates $ \varphi $. Therefore, by creating an input filter (shown in Fig. \ref{fig01}) based on $vranges$, it is possible to filter the violating input value.

\section{Formal Verification of Decision Tree Ensemble Model}\label{sec_fv}

The procedure of $formal\_verification$ in Algorithm \ref{al1} is shown in detail hereafter. For any decision tree $t_{i} \in T$ that constitutes DTEM $M$, a set of all paths extractable from $t_{i}$ is denoted by $P_{i}$. As stated in Section \ref{subsec_tree}, since a decision tree is assumed to be acyclic, the elements of $P_{i}$ are paths of finite length. Arbitrary path $p \in P_{i}$ can be represented by a sequence of nodes, $n_{1}^{p}, n_{2}^{p}, ..., n_{d}^{p}, n_{d+1}^{p}$. Among these nodes, $n_{1}^{p}, ..., n_{d}^{p}$ represent decision nodes, and $n_{d+1}^{p}$ represents a leaf node. Since $t_{i}$ has one or more decision nodes, $1 \leq d$ holds. Here, the specification of an arbitrary path $p$ of $t_{i}$ is defined as $f_{i}^{p}$ by using functions $attr$, $tv$, and $dv$ (defined in Section \ref{subsec_tree}) as follows:
\begin{multline}
f_{i}^{p} \overset{\mathrm{def}}{=} \left( \bigwedge_{j=1}^{d} attr(n_{j}^{p})(x) = tv( \langle n_{j}^{p}, n_{j+1}^{p} \rangle ) \right) \\
\rightarrow ( y_{i} = dv(n_{d+1}^{p}) )
\label{encoding_path}
\end{multline}
The argument $ \langle n_{j}^{p}, n_{j+1}^{p} \rangle $ of function $tv$ represents an edge between nodes $n_{j}^{p}$ and $n_{j+1}^{p}$. By using $f_{i}^{p}$, the specification $F_{i}$ of $t_{i}$ is expressed as follows:
\begin{equation}
F_{i} \overset{\mathrm{def}}{=} \bigwedge_{p \in P_{i}} f_{i}^{p}
\end{equation}
As explained in Section \ref{subsec_dtem}, the output value $y$ of $M$ is calculated by $ensem$ from decision values $y_{1}, ..., y_{i}, ..., y_{card(T)}$. Therefore, the specification of $M$ is expressed by $F_{M}$ by using $ensem$ and $f_{i}^{p}$ as follows:
\begin{multline}
F_{M} \overset{\mathrm{def}}{=} \left( \bigwedge_{i=1}^{card(T)} F_{i} \right) \\
\land y = ensem(y_{1}, ..., y_{i}, ..., y_{card(T)})
\label{encoding_model}
\end{multline}
Furthermore, $F$ is defined by using $ \varphi $ (i.e., a property to be satisfied by $M$) and constraint $\rho$ (which indicates that $x$ is included in domain $X$) as follows:
\begin{equation}
F \overset{\mathrm{def}}{=} F_{M} \land \lnot \varphi \land \rho
\label{veri_formula}
\end{equation}
$F$ is input into the SMT solver to determine its satisfiability. If $F$ is unsatisfiable, it is guaranteed that no input value satisfies $ \varphi $ in the range constrained by $ \rho $. If $x$ and $y$ satisfying $F$ are detected, the detected $x$ is returned as the violating input value, $ce$. It is thus possible to verify whether $M$ satisfies $ \varphi $ by encoding $M$ to formula $F$ and solving the satisfiability problem by using the SMT solver. 

\section{Violation Range Extraction}\label{sec_extract}
\subsection{Overview}
\begin{figure}[htb]
\begin{center}
\scalebox{0.55}{\includegraphics[bb=0 0 451 327]{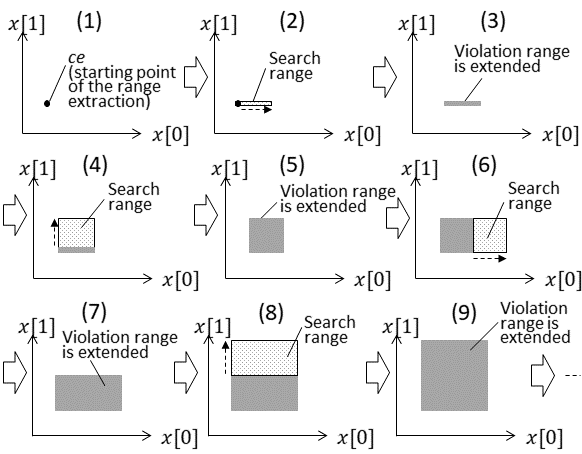}}
\end{center}
\caption{Violation range extraction by $range\_extraction$}
\label{fig02}
\end{figure}

As for $range\_extraction$, a violating input value is searched in the range around violating input value $ce$ as a starting point by using $formal\_verification$. If a violating input value is detected, the range is extracted as the violation range. Here, to promote intuitive understanding, it is assumed that variable $x$ representing the input value of $M$ is represented by a two-dimensional vector, namely, $[x[0]], x[1]]$. The method for extracting the violation range by $range\_extraction$ is outlined in Fig. \ref{fig02}. 
The violation range is extended by alternately increasing the values of $x[0]$ and $x[1]$. In step (1), the violating input value obtained on line \ref{veri} of Algorithm \ref{al1} is set as the initial value of the violation range. Next, in step (2), a search range is set from the initial value to upper direction of $x[0]$, and whether a violating input value exists within that search range is verified by the method described in Section \ref{sec_fv}. If a violating input value is detected in the search range, as shown in step (3), the search range is imported into the violation range. In step (4), the search range is set to upper direction of $x[1]$. From then on, the same procedure is repeated to extend the violation range. Although omitted in Fig. \ref{fig02}, in a similar manner as described above, the violation range is also extended to lower directions of $x[0]$ and $x[1]$.

\begin{figure}[htb]
\begin{center}
\scalebox{0.55}{\includegraphics[bb=0 0 451 327]{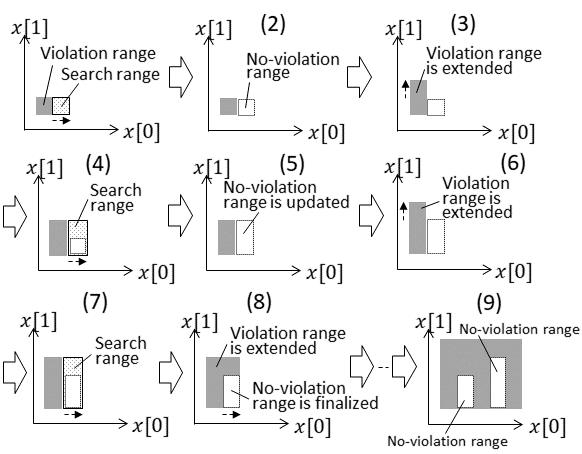}}
\end{center}
\caption{No-violation range extraction by $range\_extraction$}
\label{fig03}
\end{figure}

As for $range\_extraction$, at the same time the violation range is extracted, no-violation ranges existing in that range are extracted. The method of extracting no-violation ranges is outlined in Fig. \ref{fig03}.
In step (1) of Fig. \ref{fig03}, the search range is set in the $x[0]$ upper direction. Here, it is assumed that no violating input value exists in the search range. In that case, the no-violation range is created as shown in step (2). After the violation range is extended in the $x[1]$ upper direction in step (3), the search range is reset in the $x[0]$ direction in step (4). If a violating input value still does not exist in the search range, the no-violation range set in step (2) is extended by step (5). It is assumed that the violation range can be extended in steps (6), (7), and (8). In that case, the extension of the no-violation range is ended. Accordingly, when no violating input value exists in the search range, the search range is extracted as the no-violation range. As shown in step (9), a plurality of no-violation ranges may be created. One of the features of the proposed method is to extract this no-violation range. The extracted no-violation range is the utilized in $range\_division$ described in Section \ref{sec_divide}.

Regarding the proposed method, as a strategy for extending the violation range, alternately increasing the values of $x[0]$ and $x[1]$ is adopted. Other strategies, however, may be considered to extend the violation range. For example, a strategy by which the values of $x[0]$ and $x[1]$ are increased simultaneously can be considered. Strategies for extending the violation range are discussed in Section \ref{sec_discuss}.

\subsection{Algorithm}
The procedure of $range\_extraction$ is shown in detail in Algorithm \ref{al2}.
\begin{algorithm}[htb]
\caption{Algorithm of $range\_extraction$}
\label{al2}
\begin{algorithmic}[1]
\Require $ce$, $M$, $ \varphi $, $\rho$, $r_{a}$, $X$
\Ensure $vio$, $novios$
\newline

\State $vio[lower]$ $\leftarrow$ $ce$ \label{2_1}
\State $vio[upper]$ $\leftarrow$ $ce$ \label{2_8}
\State $novios$ $\leftarrow$ $\emptyset$
\State $tmp\_nv[lower]$ $\leftarrow$ [None] * $s$
\State $tmp\_nv[upper]$ $\leftarrow$ [None] * $s$ 
\State $cont\_flag$ $\leftarrow$ True \label{2_2}

\While{$cont\_flag$ = True} \label{2_3}
  \State $cont\_flag$ $\leftarrow$ False \label{2_4}
  \For{each $k$ in $\{ 0, ..., s-1 \}$} \label{2_5}
    \State $rtn\_upper$ $\leftarrow$ EXPAND($upper$, $k$, $vio$, $tmp\_nv$, $\rho$, $M$, $ \varphi $, $X$)\label{2_6}
    \State $rtn\_lower$ $\leftarrow$ EXPAND($lower$, $k$, $vio$, $tmp\_nv$, $\rho$, $M$, $ \varphi $, $X$)\label{2_7}
    \If{$rtn\_upper$ = True $\lor$ $rtn\_lower$ = True}
      \State $cont\_flag$ $\leftarrow$ True
    \EndIf
  \EndFor 
\EndWhile
\State \Return $vio$, $novios$
\newline
\Procedure{expand}{$dir$, $k$, $vio$, $tmp\_nv$, $\rho$, $M$, $ \varphi $, $X$}
    \State $sr$ $\leftarrow$ $vio$ \label{2_13}
    \If{$dir$ = $upper$}
        \State $sr[dir][k]$ $\leftarrow$ $vio[dir][k]$ + $mgn(r_{a})[k]$ \label{2_15}
    \Else \ // $dir$ = $lower$
        \State $sr[dir][k]$ $\leftarrow$ $vio[dir][k]$ - $mgn(r_{a})[k]$
    \EndIf
    \State $sr$[$opposite$($dir$)][$k$] $\leftarrow$ $vio[dir][k]$ \label{2_16}
    
    \State $sr$ $\leftarrow$ $intersection$($X$, $sr$) \label{2_25}
    \State $\rho'$ $\leftarrow$ $\rho$ $\land$ $input\_is\_within$($M$, $sr$) \label{2_18}
    \State $ce'$ $\leftarrow$ $formal\_verification$($M$, $ \varphi $, $\rho'$) \label{2_19}
    \If{$ce'$ $\neq$ None}
      \State $vio[dir][k]$ $\leftarrow$ $sr[dir][k]$ \label{2_20}
      \If{$tmp\_nv[dir][k]$ $\neq$ None} \label{2_21}
        \State $novios$.push($tmp\_nv[dir][k]$) \label{2_22}
        \State $tmp\_nv[dir][k]$ $\leftarrow$ None \label{2_23}
      \EndIf 
      \State \Return True
    \Else
      \State $tmp\_nv[dir][k]$ $\leftarrow$ $sr$ \label{2_24}
      \State \Return False
    \EndIf
\EndProcedure
\end{algorithmic}
\end{algorithm}
Algorithm \ref{al2} returns $vio$, which represents a violation range for $ \varphi $ , and $novios$, which represents a range in which no violating input value exists. The violation range can be defined by the lower-limit values and the upper-limit values of variables $x[0], ..., x[s-1]$. Therefore, it is supposed that violation range $vio$ is composed of $vio[lower]$ and $vio[upper]$, where $lower$ and $upper$ represent $0$ and $1$ respectively. Here, the lower-limit value of variables $x[k]$ ($0 \leq k \leq s-1$) is represented by $vio[lower][k]$. Similarly, the upper-limit value is represented by $vio[upper][k]$.

First, on lines \ref{2_1} and \ref{2_8}, $vio$ is initialized with $ce$. On line \ref{2_3}, whether $vio$ has been extended in the previous loop is checked, and if $vio$ has been extended, a further extension is tried. On line \ref{2_5}, the variable for which the range is to be expanded, $x[k]$, is selected. For the selected $x[k]$ here, line \ref{2_6} attempts to expand the upper limit of the variable, and line \ref{2_7} tries to expand the lower limit of the variable. If neither the upper limit nor the lower limit can be expanded for all variables $x[0], ..., x[s-1]$, $cont\_flag$ = False is returned, and the procedure ends.

The upper and lower limits are extended on lines \ref{2_6} and \ref{2_7} by EXPAND. The procedure of EXPAND is described below with the case that the upper limit is expanded taken as an example. The parameter $dir$ of EXPAND represents the direction of expansion. That is, in this example, $upper$ is passed as an argument. On line \ref{2_13}, variable $sr$ is initialized with $vio$. where $sr$ represents the search range. On lines \ref{2_15} and \ref{2_16}, $sr$ for variable $x[k]$ is updated. $opposite$ is a function that returns the direction opposite to the extension direction indicated by $dir$. $mgn$ is a function that accepts parameter $r_{a}$ as an argument and returns a vector of the same length as $x$. Vector $mgn(r_{a})$ is used to set the search range of $x[k]$ to the range $vio[upper][k] < x[k] \leq (vio[upper][k] + mgn(r_{a})[k])$. 
Because the search range of variable $x[k'] (k' \neq k)$ is not updated, it is given as $vio[lower][k'] \leq x[k'] \leq vio[upper][k']$. As a result, $sr$ is created on the upper bound of $vio$ in the $x[k]$ direction.

When $sr$ is defined, the upper-limit or lower-limit value of $sr$ is sometimes not included in the search range, that is, the search range is defined by using ``<" or ``>."  Accordingly, it is necessary to hold that information about with-equal or without-equal. In this paper, to simplify explanation, it is omitted how to keep that information in $sr$. The function $intersection$ on line \ref{2_25} redefines the common range of $X$ and $sr$ as a new $sr$. As the violation range is extended, $sr$ is eventually created outside $X$, and then their common ranges become empty. In this way, the termination of the procedure is assured.

On line \ref{2_18}, $ \rho' $ is created by conjunctively appending the constraint ``the input value for $M$ is included in $sr$" to $ \rho $. Line \ref{2_19} verifies whether $M$ satisfies $ \varphi $ under constraint $ \rho' $. As a result of verification, if the violating input value is detected within $sr$, the upper limit of the violation range, $vio[upper]$, is extended to $sr[upper]$ (line \ref{2_20}).

Lines \ref{2_21}, \ref{2_23}, and \ref{2_24} create an element of $novios$. If no violating input value is detected in the search range as a result of the verification on line \ref{2_19}, the search range is saved in $tmp\_nv[upper][k]$ (line \ref{2_24}). $tmp\_nv[upper][k]$ stores a tentative no-violation range in the $x[k]$ $upper$ direction. The range stored in $tmp\_nv[upper][k]$ will be extended as long as a violating input value is not detected in the $x[k]$ $upper$ direction (line \ref{2_24}). If a violating input value is detected in that direction, the latest range stored in $tmp\_nv[upper][k]$ is finalized as a no-violation range and added to $novios$ (line \ref{2_22}). 

\section{Violation Range Division}\label{sec_divide}
\subsection{Overview}
\begin{figure}[htb]
\begin{center}
\scalebox{0.55}{\includegraphics[bb=0 0 422 112]{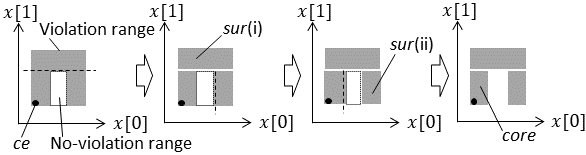}}
\end{center}
\caption{Violation range divided by $range\_division$}
\label{fig04}
\end{figure}

As for $range\_division$, violation range $core$ extracted by $range\_extraction$ is divided. The division is based on a no-violation range, which is an element of $novios$. Among the divided ranges, the range including violating input value $ce$ (which is the starting point of the range extension) is set as a new $core$, and the other ranges are called $sur$s. If $sur$ includes a violating input value, it is added to $surrds$. Since it is possible to repeatedly divide $core$, the number of ranges in $surrds$ may be increased with each division. For example, as for the violation range shown in step (9) in Fig. \ref{fig03}, the method of $range\_division$ is outlined in Fig. \ref{fig04}
In Fig. \ref{fig04}, first, the violation range is divided at the upper-limit value of the no-violation range on the $x[1]$ axis. Next, the range is divided at the upper-limit value of the no-violation range on the $x[0]$ axis. Finally, the range is further divided by using the lower-limit value of the no-violation range on the $x[0]$ axis. As a result this division, the violation range is divided into three ranges: $sur$(i), $sur$(ii), and $core$.

It is clear that there exits a violating input value in $core$ because $core$ includes violating input value $ce$ at least. $sur$(i) also includes another violating input values in this case since the violation range was extended as shown in step (8) in Fig. \ref{fig03}. However, $sur$(ii) may not include a violating input value. Whether $sur$(ii) include a violating input value can be checked by using $formal\_verification$. If $sur$(ii) does not include a violating input value, it is removed from the violation range.

This method aims to narrow the violation range by dividing it and verifying whether outer ranges $sur$s created by the division include violating input values. 
This is based on the assumption that outer ranges of the no-violation range are also likely not to include a violating input value. Outer ranges are ``adjacent to" the no-violation range and also ``beyond" the no-violation range from the starting point of the range extension. Accordingly, input values in outer ranges are assumed to be ``similar to" the input values in the no-violation range and also ``more different" from violating input values around the starting point than the input values in the no-violation range. Therefore, we consider that outer ranges created by the division on the basis of the no-violation range are likely to include no violating input value.

\begin{figure}[htb]
\begin{center}
\scalebox{0.55}{\includegraphics[bb=0 0 422 112]{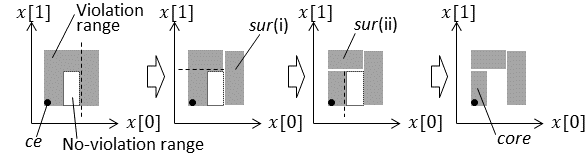}}
\end{center}
\caption{Another order of dividing based on no-violation range}
\label{fig05}
\end{figure}

When dividing the violation range on the basis of the no-violation range, the order of the division can be changed.
For example, the dividing order shown in Fig. \ref{fig05} can be considered. In this case, the division is performed in the following order: at the upper-limit value of the no-violating range on the $x[0]$ axis, at the upper-limit value on the $x[1]$ axis, and at the lower-limit value on the $x[0]$ axis. 

It is said that the dividing order that makes the violation range narrower is more effective. However, the extent to which the violation range can be narrowed by a certain dividing order depends on the DTEM and property to be verified. It is therefore not known which dividing order is effective unless the division is actually performed. Accordingly, as for the proposed method, multiple divisions are tried at random, and the order that makes the violation range the narrowest is adopted. 

The violation range is divided by the surface of the no-violation range (corresponding to the dotted lines in Fig. \ref{fig04} and Fig. \ref{fig05}). Since the no-violation range is an $s$-dimensional hyperrectangle, the number of dividing planes is $2s$. Therefore, the number of ways to divide the violation range is $(2s)!$
As for the proposed method, dividing orders, whose number equals parameter $r_{c}$, are randomly selected and the results of the divisions are compared according to the sum of the hypervolumes of the ranges in $surrds$. The dividing order that minimizes the sum of the hypervolumes is adopted

\subsection{Algorithm}\label{divide_al}
The procedure of $range\_division$ is shown in details in Algorithm \ref{al3}. 
\begin{algorithm}[htb]
\caption{Algorithm of $range\_division$}
\label{al3}
\begin{algorithmic}[1]
\Require $core$, $iv$, $\rho$, $r_{c}$, $M$, $ \varphi $
\Ensure $core$, $surrds$
\newline

\State $planes$ $\leftarrow$ $calc\_planes$($iv$) \label{3_1}
\State $orders$ $\leftarrow$ $permutate$($plane$, $r_{c}$) \label{3_2}
\State $results$ $\leftarrow$ $\emptyset$
\For{each $od$ in $orders$} \label{3_3}
  \State $core$ $\leftarrow$ $core$ \label{3_10}
  \While{$od$ $\neq$ $\emptyset$}
    \State $plane$ $\leftarrow$ $od$.pop()\label{3_4}
    \State $core$, $sur$ $\leftarrow$ $slice\_core$($core$, $plane$) \label{3_5}
    \State $\rho'$ $\leftarrow$ $\rho$ $\land$ $input\_is\_within$($M$, $sur$)
    \State $ce'$ $\leftarrow$ $formal\_verification$($M$, $ \varphi $, $\rho'$) \label{3_6}
    \If{$ce'$ $\neq$ None}
      \State $surrds$ $\leftarrow$ $surrds$ $\cup$ $sur$ \label{3_7}
    \EndIf
  \EndWhile
  \State $volume$ $ \leftarrow $ $calc\_volume\_sum$($core$, $surrds$) \label{3_8}
  \State $results$ $\leftarrow$ $results$ $\cup$ $\{((core, surrds), volume)\}$ \label{3_9}
\EndFor
\State $(core, surrds)$ $\leftarrow$ $min\_volume$($results$) \label{3_11}
\State \Return $core$, $surrds$
\end{algorithmic}
\end{algorithm}
In Algorithm \ref{al3}, violation range $core$ is divided on the basis of no-violation range $iv$. Then, as a result of division, a new $core$ and $surrds$ are returned. On line \ref{3_1}, $calc\_planes$ extracts the ($s$-1)-dimensional hyperplanes constituting hyperrectangle $iv$, and creates the set $planes$. The number of elements in $planes$ is $2s$. On line \ref{3_2}, $permutate$ creates as many arbitrary dividing orders using elements of $planes$ as $r_{c}$, and the set of the created dividing orders is assined to $orders$. Line \ref{3_3} picks out arbitrary element $od$ of $orders$. $od$ is an array (with length of $2s$) having the hyperplanes constituting $iv$ as elements. On line \ref{3_4}, the top element of $od$ is assined to $plane$. By $slice\_core$ on line \ref{3_5}, $core$ is divided with $plane$ as the dividing plane. Of the two ranges created by the division, the range that includes $ce$ is taken as the new $core$, and the other is taken as $sur$. $formal\_verification$ on line \ref{3_6} verifies whether the violating input value is included in the range of that $sur$. If a violating input value exists within the range of $sur$, $sur$ is added to $surrds$ (line \ref{3_7}). The division with $plane$ is repeated for each element of $od$. On line \ref{3_8}, $calc\_volume\_sum$ is used to calculate the sum of the hypervolumes of the violation range represented by the elements of $core$ and $surrds$. 
The procedures from lines \ref{3_10} to \ref{3_9} are repeated according to the number of elements of $orders$, that is, $r_{c}$ times. In this manner, for each dividing order randomly created on line \ref{3_2}, $core$ and $surrds$, which are division results, and $volume$, which is the sum of the hypervolumes, are obtained. On line \ref{3_11}, the result of division that gives the smallest $volume$ is extracted by function $min\_volume$, and is used as the final return value. 

In this way, the violation range is divided by $range\_division$ on the basis of the no-violation range created by $range\_extraction$. It should be noted that $range\_division$ is not executed if no-violation range is created by $range\_extraction$.

\section{Case Study}\label{sec_casestudy}
\subsection{Setup}\label{sec_setup}
As a subject of the case study, DTEM $M$ is created by using a dataset of house prices \footnote{https://www.kaggle.com/harlfoxem/housesalesprediction} as a training data set. XGBoost \cite{xgboost} was adopted to implement $M$. $M$ receives a $7$-dimensional vector, $x = [x[0], ..., x[6]]$, as an input value, and returns a house price as output value $y$, where $x[0], ..., x[6]$ are, respectively, grade of house, condition of house, number of bedrooms, size of living room, size of parking space, size of ground floor, and size of basement. The number of decision trees constituting $M$ is $100$, and the maximum depth of each decision tree is $3$. As properties to be verified, $\varphi_{1}$, $\varphi_{2}$, and $\varphi_{3}$ are defined as follows:
\begin{eqnarray}
\varphi_{1} &\overset{\mathrm{def}}{=}& x[0] \geq 7000 \Rightarrow y \geq 500000 \\
\varphi_{2} &\overset{\mathrm{def}}{=}& y > 50000 \\
\varphi_{3} &\overset{\mathrm{def}}{=}& y < 10000000 
\label{def_property}
\end{eqnarray}
Here, $x[0]$ in $\varphi_{1}$ represents size of living room. The larger the living room, the higher the price. Therefore, it is defined as $\varphi_{1}$ that if the size of the living room is 7000 or more, the price is 500,000 or more. Moreover, the price output by $M$ is expected to be realistic in regard to a house price. Accordingly, $\varphi_{2}$ is defined as expressing that the price is higher than 50,000. Similarly, $\varphi_{2}$ is taken as the property that the price is less than 10,000,000.

Next, domain $X$ of $x$ is defined. In this case study, $X$ is defined on the basis of the maximum and minimum values of the training dataset. For each variable $x[k]$ ($0 \leq k \leq 6$), the maximum value included in the training dataset is represented as $max_{k}$, and the minimum value is similarly represented as $max_{k}$. These values are used to define $X$ as follows:
\begin{multline}
X \overset{\mathrm{def}}{=} \{ [x[0], ..., x[k], ..., x[s-1]] \\
\mid \forall k \cdot min_{k} \leq x[k] \leq max_{k} \}
\label{def_X}
\end{multline}

Function $mgn$ (which determines the search range in $range\_extraction$) is implemented with parameter $r_{a}$ as an argument as follows:
\begin{multline}
\label{def_param_a}
mgn(r_{a}) \overset{\mathrm{def}}{=} [m[0], ..., m[k], ..., m[s-1]] \mbox{, such that } \\
\left\{ \begin{array}{ll}
\forall k \cdot m[k] = max_{k} - min_{k}) / r_{a} \ \ \ \ \ \ \ \ \ (x[k] \in \mathbb{R}) \\
\forall k \cdot m[k] = ceil((max_{k} - min_{k}) / r_{a}) \ (x[k] \in \mathbb{Z}) \\
\end{array} \right.
\end{multline}
Here, $mgn(r_{a})[k]$ is created on the basis of the width of the domain $X$ (i.e., the difference between $max_{k}$ and $min_{k}$), where function $ceil$ rounds up a real number to an integer. In this case study, $r_{a} = 100$ is supposed. $continue\_division$ shown in Algorithm \ref{al1} is implemented so as to return ``True" if the hypervolume of $core$ is $r_{b}$ percent or more of the hypervolume of $X$. In this case study, $r_{b} = 10$ is also supposed. Furthermore, parameter $r_{c}$ for determining the number of elements of $orders$ in $range\_division$ is taken as $r_{c} = 10$.

The proposed method was implemented in Python \footnote{Available on https://github.com/hitachi-rd-yokohama\\ /deep\_saucer/tree/master/xgb\_encoding}. The Z3 Theorem Prover \cite{z3} was used as a SMT solver in $formal\_verification$. Moreover, this case study was performed on a Windows 10$^{\textregistered}$ \ PC equipped with two Intel$^{\textregistered}$ Core\texttrademark \ i7-8700 3.2-GHz processor with 6 cores, 16-GB memory.

\subsection{Results and Evaluation}\label{sec_results}
The results of applying the proposed method are listed in Table \ref{table01}. The violation range detected by the proposed method is shown in column (b) for each property shown in column (a). In column (b), the violation range before division by $range\_division$ and the violation range after division are shown. In column (c), the hypervolume of each violation range is shown. Column (d) shows a value obtained by dividing the sum of hypervolumes of the violation range after division by the sum of hypervolumes of the violation range before division. This value is used to evaluate the effect of narrowing the violation range by $range\_division$. Column (e) shows the time elapsed to execute the proposed method. Column (f) shows the time taken to execute the SMT solver as part of the method. Column (g) shows the number of times of the SMT solver is executed (i.e., number of calls), and column (h) shows the result of dividing (f) by (g), that is, the average execution time per run of the SMT solver.
\begin{table*}[htb]
\begin{center}
\scalebox{0.70}{\includegraphics[bb=0 0 1001 307]{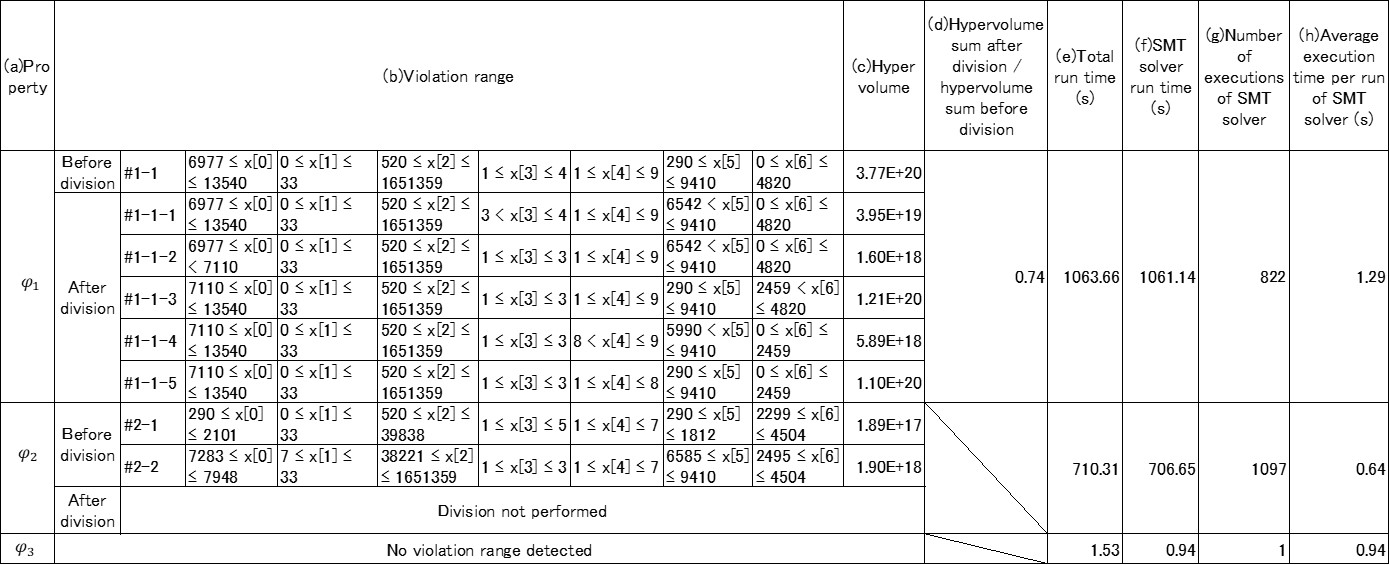}}
\end{center}
\caption{Violation ranges extracted by the proposed method}
\label{table01}
\end{table*}
As a result of executing $formal\_verification$ for $\varphi_{1}$, a violating input value was detected. Then, starting from the detected violating input value, $range\_extraction$ is executed, and the violation range shown in \#1-1 is extracted. Moreover, by dividing this violation range by $range\_division$, the violation ranges shown in \#1-1-1 to \#1-1-5 were obtained. This division reduced the hypervolume of the violation range to about 74 percent of the hypervolume before the division.

Similarly, with respect to $\varphi_{2}$, the violation ranges shown in \#2-1 and \#2-2 were extracted by $range\_extraction$. In this case, 
the division by $range\_division$ was not performed because the hypervolume of each violation range was less than $r_{b} = 10$ \%
 of the hypervolume of $X$. However, it is confirmed that when $r_{b}$ is $0.1$, $range\_division$ reduces the hypervolume of the violation range to about 33\%
 of the violation range before the division. As for $\varphi_{3}$, no violating input value was detected. That is, it was confirmed that $M$ satisfies $\varphi_{3}$.

The above results demonstrate that the DTEM can be formally verified by the method described as $formal\_verification$, that is, encoding the DTEM in a formula and solving it with an SMT solver. Moreover, it was confirmed that the violation range can be extracted by $range\_extraction$. It was also confirmed that using $range\_division$ makes it possible to narrow the violation range. It can be therefore be concluded that the feasibility of the proposed method was demonstrated.

Next, we evaluate the scalability of the proposed method. If total execution time and execution time of the SMT solver are focused on, it becomes clear that 
the time taken to execute the SMT solver occupies most of the total execution time. In other words, the execution time of the proposed method is considered to be largely dependent on the execution time of the SMT solver. As a factor that affects the execution time of the SMT solver, for example, the number and depth of the decision trees constituting $M$ can be considered. However, various heuristics are implemented and black-boxed in the SMT solver, so it is difficult to predict how much these factors affect the execution time of the SMT solver. Accordingly, in this study, we change the value of the factor considered to affect the execution time of the SMT solver and measure the execution time. The following experiment uses the same settings as described in Section \ref{sec_setup} unless otherwise stated. Moreover, a practically acceptable execution time is assumed as 24 hours, and the execution is aborted if it exceeds 24 hours.

\subsubsection{(1) Number of decision trees}
The number of decision trees constituting $M$ is represented by $n\_est$. In proportion to the increase of $n\_est$, the number of paths of $M$ increases. The length of formula $F_{M}$ increases in proportion to the number of paths of $M$. That is, with the increase of $n\_est$, the length of the formula $F_{M}$ increases by $O(n\_est)$. Therefore, it is considered that the execution time of the SMT solver tends to increase with the increase of $n\_est$. The execution times of the proposed method when the value of $n\_est$ was changed are listed in Table \ref{table02}.
\begin{table*}[htb]
\begin{center}
\scalebox{0.70}{\includegraphics[bb=0 0 932 350]{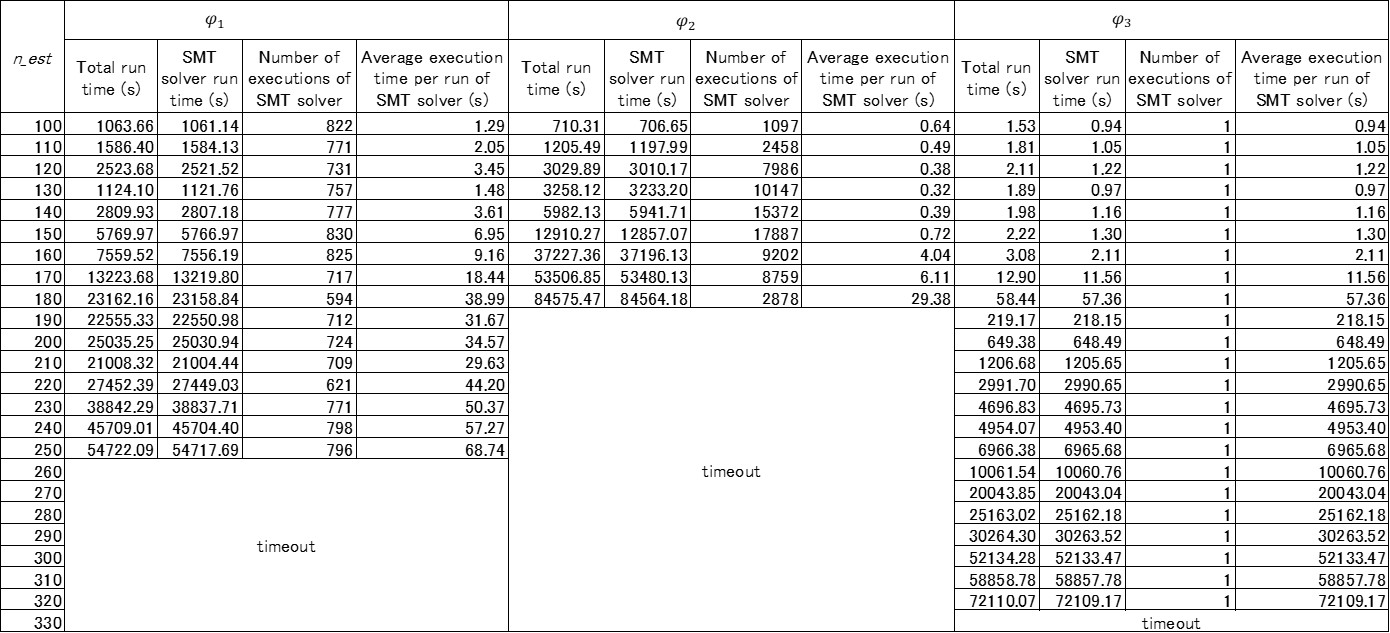}}
\end{center}
\caption{Execution time when $n\_est$ is changed}
\label{table02}
\end{table*}
The results in Table \ref{table02} show that the execution time of the SMT solver tends to increase as $n\_est$ increases. As for the verification of $\varphi_{2}$, timeout occurs when $n\_est \geq 190$. If the number of executions of the SMT solver is focused on, it turns out that the SMT solver was executed frequently in the case of $\varphi_{2}$.
In such a case, changing parameters $r_{a}$, $r_{b}$, and $r_{c}$ may reduce the number of executions of the SMT solver and, thereby, shorten the execution time.

Parameter $r_{a}$ determines the size of the search range in $range\_extraction$. In the implementation shown in Equation \ref{def_param_a}, the smaller the value of $r_{a}$ is , the larger the search range is. Therefore, the number of times to call $formal\_verification$ can be reduced if the value of $r_{a}$ is changed smaller. Note that, in that case, it is highly likely that the extracted violation range is wider than before decreasing $r_{a}$. That is, the extracted violation range includes more input values that do not violate the property.
$r_{b}$ is referred to as a reference value for determining whether or not to execute $range\_division$. In the implementation shown in Section \ref{sec_setup}, the larger the value of $r_{b}$ is, the lower the possibility of dividing the violation range is. Thus, the number of times to call $formal\_verification$ is reduced if the value of $r_{b}$ is changed larger. In that case, however, the possibility of narrowing the violation range is reduced. $r_{c}$ determines how many dividing orders are tried in $range\_division$. Therefore, if the value of $r_{c}$ is changed smaller, the number of times to call $formal\_verification$ decreases. We changed the values of these parameters to $r_{a} = 20$, $r_{b} = 30$, and $r_{c} = 5$ and then retried the verification of $\varphi_{2}$. Execution times before and after changing the parameters are listed in Table \ref{table03}.
\begin{table*}[htb]
\begin{center}
\scalebox{0.70}{\includegraphics[bb=0 0 834 300]{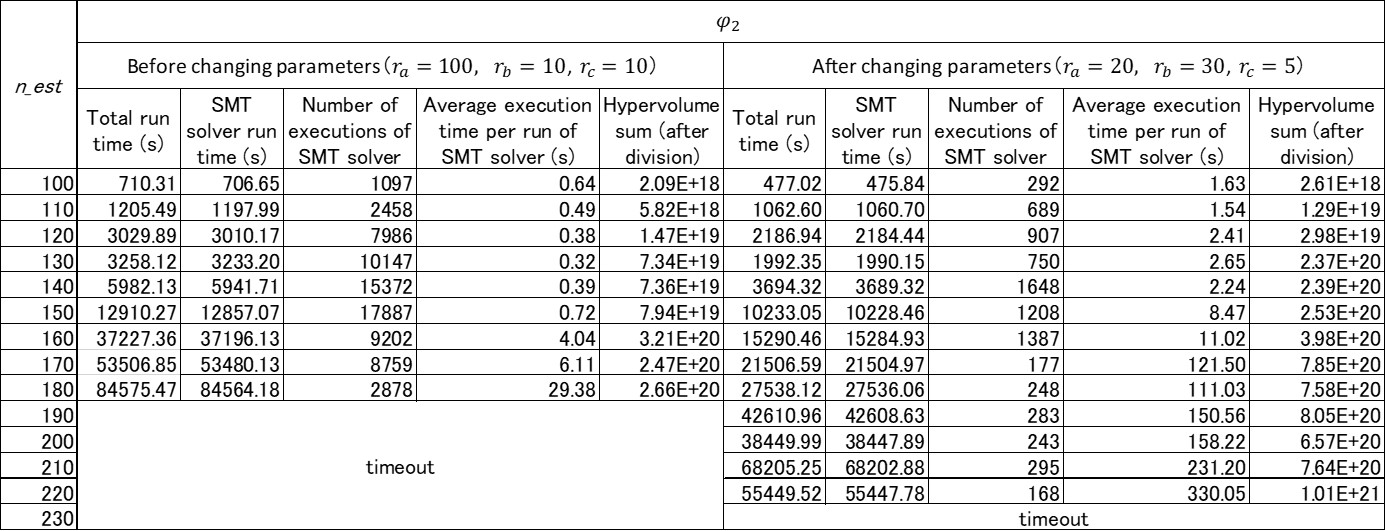}}
\end{center}
\caption{Execution times before and after changing parameters}
\label{table03}
\end{table*}
As shown in Table \ref{table03}, by changing the values of $r_{a}$, $r_{b}$, and $r_{c}$, even in the case of $190 \leq n\_est \leq 220$, execution of the SMT solver could be completed without incurring the timeout. This demonstrates that when execution cannot be completed within a practical time due to the number of executions of the SMT solver, the execution time can be shortened by adjusting the values of these parameters. It should, however, be noted that when these parameters are adjusted to reduce the execution time, the extracted violation range may become wider. In fact, 
from Table \ref{table03}, it turns out that the total volume of the violation range is increased by changing the value of parameters. On the contrary, it may be possible to narrow the violation range more strictly by changing these parameters if it is acceptable that the execution time become longer. That is, by changing these parameters, it is possible to adjust the balance between the fineness of the result and the execution time.

\subsubsection{(2) Depth of decision trees}
The maximum depth of a decision tree is taken as $max\_d$. When the depth of a decision tree increases by 1, the number of paths extracted from one decision tree is doubled. (Moreover, the length of each path also increases in proportion to $max\_d$.) Therefore, the length of formula $F_{M}$ is of $O(2^{max\_d})$. 
Accordingly, if $max\_d$ increases, it is considered that the execution time of the SMT solver tends to increase. The execution times of the proposed method when the value of $max\_d$ was changed are listed in Table \ref{table04}.
\begin{table*}[htb]
\begin{center}
\scalebox{0.70}{\includegraphics[bb=0 0 884 200]{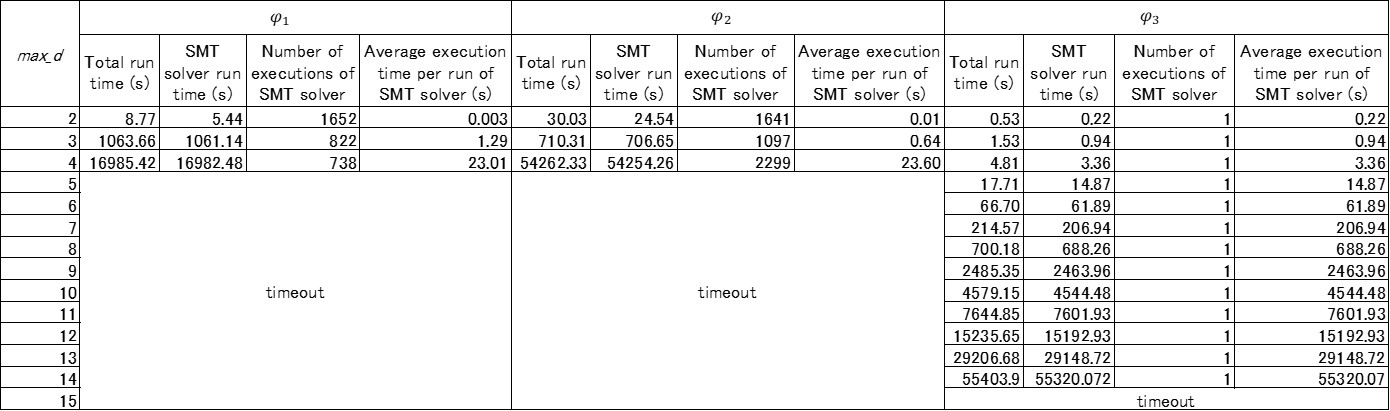}}
\end{center}
\caption{Execution times in case of changing $max\_d$}
\label{table04}
\end{table*}
It is clear from the results in Table \ref{table04} that the average execution time of the SMT solver increases as $max\_d$ increases. In particular, as for the verification of $\varphi_{1}$ and $\varphi_{2}$, the execution time increases by 10 to 100 times each time $max\_d$ increases by one. When time-out is invalidated and $\varphi_{1}$ is verified with $max\_d$ = 5, it takes about 70 hours to complete execution, and the average execution time of SMT solver is about 46 seconds. From these results, it can be said that $max\_d$ has a big influence on the execution time of the proposed method.

\subsubsection{(3) Dimension of input value}
The loop in line \ref{2_5} of $range\_extraction$ in which $formal\_verification$ is called is executed by the number of dimension $s$. In addition, as $s$ increases, the number of variables appearing in $F_{M}$ also increases, that is, $F_{M}$ becomes more complicated. For the reasons above, it is considered that the execution time of the SMT solver tends to increase with increasing $s$. In the house-price dataset used in this study, we can use up to 18 attributes as elements of $x$.
Therefore, these attributes was used to change dimension $s$ of $x$ in the range of $ 2 \leq s \leq 18$. The execution times of the proposed method when the value of $s$ was changed are listed in Table \ref{table05}.
\begin{table*}[htb]
\begin{center}
\scalebox{0.70}{\includegraphics[bb=0 0 884 270]{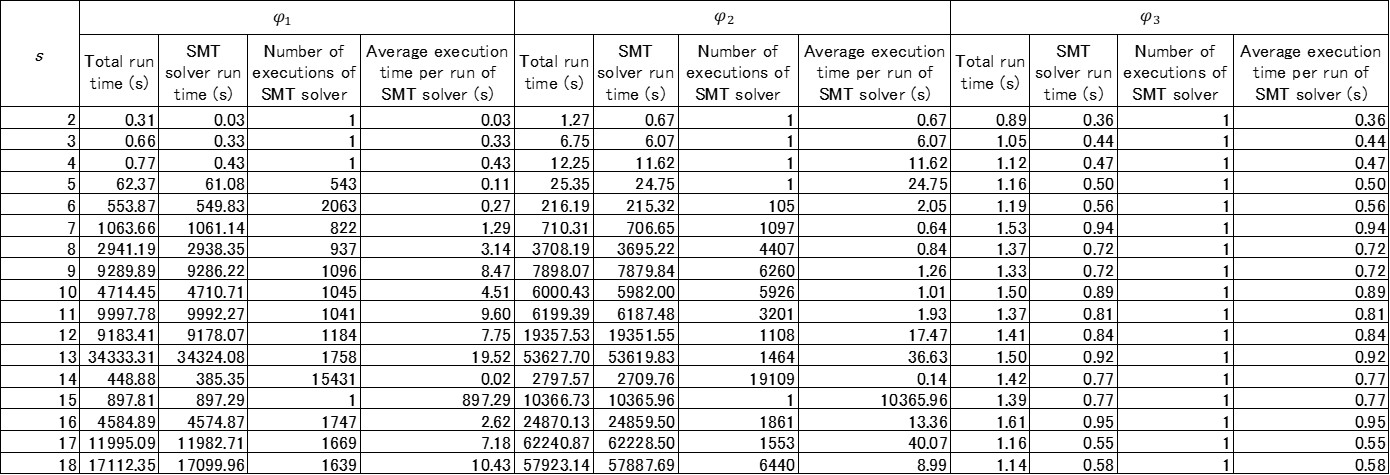}}
\end{center}
\caption{Execution times in case of changing dimension $s$}
\label{table05}
\end{table*}
It is clear from the results in Table \ref{table05} that execution time tends to increases as $s$ increases. However, the execution time sometimes decreases though $s$ increases, for example, in the verificaiton of $\varphi_{1}$, $s = 14$. In other cases, execution time increases rapidly as when $s = 13$ in the verification of $\varphi_{1}$. Therefore, the correlation of $s$ with the execution time is considered to be weaker than that of $n\_est$ and $max\_d$ with execution time. 

From the results in Table \ref{table03}, Table \ref{table04}, and Table \ref{table05}, it can be concluded that $n\_est$, $max\_d$, and $s$ are factors that increase the execution time of the proposed method. In particular,  
$n\_est$ and $max\_d$ are important factors that determines the applicability of the proposed method. Specifically, the proposed method is practical if $n\_est$ is less than around 200 and $max\_d$ is less than 5 at least.


\section{Discussion}\label{sec_discuss}
As for the proposed method, the search range is set around the violating input value detected first, and if a violating input value exists within the search range, the search range is defined as the violation range. Furthermore, by setting the search range around the violation range, the violation range is expanded in the same manner. After the violation range is extracted, it is confirmed whether another violating input value exists outside the violation range. If another violating input value exists, the violation range is extracted in the same way starting from that violating input value. In this manner, it is assured that all violating input values fall within any violation range. Therefore, by creating the input filter shown in Fig. \ref{fig01} based on the violation ranges extracted by the propsed method, all violating input values leading to system failures can be filtered.

Moreover, as for the proposed method, the violation range can be narrowed by dividing the extracted violation range. Through a case study, the feasibility of the division of the violation range was confirmed. The extent to which the violation range can be narrowed depends on the DTEM or the property to be verified; thus, it is difficult to estimate the effectiveness of the division in advance. However, dividing the violation range is still useful to narrow the violation range on a trial basis.

In the case study described in Section \ref{sec_results}, it was shown that the number of decision trees constituting the DTEM, $n\_est$, maximum depth of the decision trees, $max\_d$, and dimension $s$ of the input value are factors that increase the execution time of the proposed method. Among those factors, $max\_d$ has the greatest influence on the execution time. On the other hand, 
the correlation of $s$ with execution time is considered to be relatively weak. Here, the length of the formula $F_{M}$ is focused on. As mentioned above, $F_{M}$ becomes longer by exponential order with increasing $max\_d$. On the contrary, the length of $F_{M}$ does not change even if $s$ increases. From these facts, it is inferred that the execution time of the proposed method strongly depends on the length of $F_{M}$. 


The purpose of the proposed method is creating the filtering condition for the input filter. In that case, the proposed method is required to complete execution within a practical time. However, even if execution of the proposed method is not completed within a practical time, the proposed method can be utilized by outputting the violation range extended up to that point when the execution was interrupted. The violation range extracted at the time execution is suspended does not include all violating input values. However, it is useful for realizing where violating input values exist around. For example, it is possible to use the input values included in the violation range as training data for additionally training the DTEM. 
Thus, the proposed method can be useful, even in the case of a large-scale DTEM whose execution is not completed within the practical time.

As described in Section \ref{sec_intro}, as a method for obtaining the condition to filter input value that violates a property, a method that outputs all the violating input values can be considered. However, for a certain violating input value, be a large number of similar violating input values might exist, and the values of some elements in them might be slightly changed. Accordingly, from the viewpoint of calculation time, it is not practical to detect all violating input values. Furthermore, as another approach, extracting the path condition \cite{symb_exec2} of the decision trees constituting DTEM is considered hereafter.

First, in the same manner as the proposed method, the violating input value is detected by verification. If the detected violating input value is input into the DTEM, the execution path of each decision tree is uniquely determined. Here, the condition of the input value to execute these paths is called ``path condition". The path condition can be extracted by executing the DTEM with the violating input value or by analyzing the decision trees.
If an input value satisfies the path condition and it is input into the DTEM, the same paths are executed as they are when the violating input value is input. Moreover, if the same paths are executed, the same output value is returned. Therefore, for all input values satisfying the path condition, the DTEM returns the same output value. Here, by assigning the output value to variable $y$ appearing in the property to be verified and negating the assigned property, a conditional expression for the input value is created. For example, let the output value be $499,999$, the conditional expression of property $\varphi_{1}$ is $ \neg (x[0] \geq 7000 \Rightarrow 499999 \geq 500000) $. If an input value satisfies the conjunction of the path condition and the conditional expression of the property, it means that the corresponding output value is $499,999$, and at that time, the property is not satisfied. Hereafter, the conjunction is called the ``violation condition."
By repeating the the above procedure, other violation conditions can be extracted. Finally, the set of the extracted violation conditions can be used as the filtering condition of the input filter.

Since the violation condition extracted by this method filters only input values that violate the property, this method creates the filtering condition more precisely than the proposed method. However, from the viewpoint of calculation time, this method is considered to be less practical than the proposed method because it is likely to extract a large number of violation conditions. For example, in the case of a DTEM composed of 100 decision trees with two paths each, there are $2^{100}$ combinations of paths, that is, about $10^{29}$ combinations. The violation condition extracted in a single verification is only one of the $10^{29}$ combinations, and it is highly likely that there will be a large number of similar violation conditions in which part of the path condition differs. In that case, since the number of verifications by the SMT solver is also enormous, this method is not practical.

\begin{figure}[htb]
\begin{center}
\scalebox{0.55}{\includegraphics[bb=0 0 422 112]{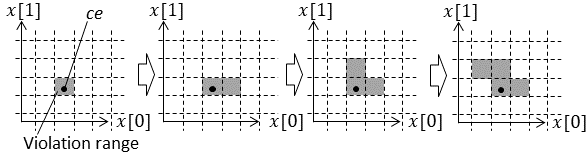}}
\end{center}
\caption{Example of extending violation range by using mesh strategy}
\label{fig06}
\end{figure}

In general, the fineness of the result and the execution time have a trade-off relationship. As for the strategy to extend the violation range, this trade-off relationship should be discussed. 
As for the proposed method, the violation range is extended in each direction of $x[k]$ ($0 \leq k \leq s-1$) in order. Hence, violation ranges are extracted in the form of hyperrectangles. The extension strategy of the proposed method is therefore called a ``hyperrectangle strategy." Here, the mesh strategy described below can be considered as a strategy for extracting the violation range more precisely than the hyperrectangle strategy. An example of extending the violation range according to mesh strategy is shown in Fig. \ref{fig06}.
As for this mesh strategy, first, the space of input values is divided into meshes. Then, the mesh including the violating input value $ce$ is taken as the initial violation range. Next, whether the violating input value exists in the mesh adjacent to the violation range is verified. If a violating input value exists in the adjacent mesh, the adjacent mesh is taken into the violation range. Then, if no violating input value exists in any mesh adjacent to the violation range, the extension of the violation range is ended. When this strategy is adopted, the number of adjacent meshes increases as the violation range is extended. For example, suppose that the violation range is an $s$-dimensional hypercube and the number of meshes on one side of the violation range is $ \alpha $. In this case, the number of adjacent meshes to the violation range is $2s * \alpha ^{s-1}$. The verification is performed for the number of adjacent meshes to extend the violation range by one round. Therefore, the amount of calculation to extend the violation range by one round increases as the violation range is extended. On the other hand, in the case of the hyperrectangle strategy, to extend the violation range by one round in the same situation, the verification is performed only $2s$ times. That is, it advantageous that the violation range can be extended with a fixed amount of calculation regardless of the size of the violation range.

\begin{figure}[htb]
\begin{center}
\scalebox{0.55}{\includegraphics[bb=0 0 378 98]{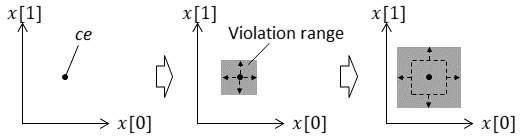}}
\end{center}
\caption{Example of extending violation range by using hypercube strategy}
\label{fig07}
\end{figure}

\begin{figure}[htb]
\begin{center}
\scalebox{0.55}{\includegraphics[bb=0 0 438 120]{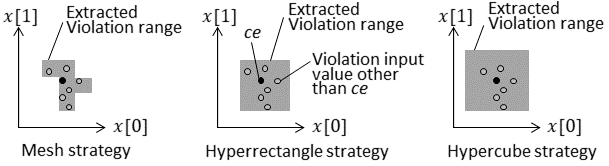}}
\end{center}
\caption{Comparison of fineness of violation range}
\label{fig08}
\end{figure}

Moreover, a strategy to increase the values of $x[1], ..., x[s-1]$ simultaneously can be considered. This strategy can extend the violation range in a shorter time than the hyperrectangle strategy. In this case, the violation range is extracted in the form of a hypercube, so this strategy is called a ``hypercube strategy." An example of extending the violation range on the basis of the hypercube strategy is shown in Fig. \ref{fig07}.
When the hypercube strategy is adopted, the number of verifications for extending the violation range by one round is one. However, the violation range extracted by the hypercube strategy is coarser than that of other strategies. An example of the extracted violation range for each strategy is shown in Fig. \ref{fig08}. The hypercube strategy is highly likely to include a range in which no violating input value exists in the violation range. Therefore, if an input filter is created based on the violation range created by the hypercube strategy, many input values satisfying the property will be filtered.

Finally, in terms of execution time, mesh strategies are considered less practical than hyperrectangle strategies. The hypercube strategy is considered to be less practical in terms of the fineness of the results. Therefore, we adopted a hyperrectangle strategy with a good balance between the fineness of the result and execution time.

As mentioned in Section \ref{sec_results}, changing the values of $r_{a}$, $r_{b}$, and $r_{c}$ makes it possible to adjust the balance between fineness of the result and execution time. These parameters should be changed by the user according to the DTEM and property to be verified, and also allowable execution time. As for finding an appropriate parameter value, the following procedure can be considered. First, the parameter values are set so that execution time is short, and it is confirmed by actually executing the procedure that the execution time can be shoter than the allowable time. After that, the parameters are gradually changed to improve the fineness of the result. This procedure improves the fineness of the results within the constraints of execution time. 

Among the proposed methods, $range\_extraction$ and $range\_division$ are applicable to machine-learning models other than a DTEM. On the contrary, $formal\_verification$ is a procedure specialized for a DTEM, so it must be to replaced with a verification method suitable for the target machine-learning model. For example, in the case of a DNN, methods for encoding and verifying the DNN in a formula have been proposed \cite{stanford_lp1}\cite{stanford_lp2}\cite{stanford_lp3}\cite{planet}\cite{oxford_smt}\cite{unified}. By replacing $formal\_verification$ in the proposed method with such a method, it is possible to extract the violation range of the DNN. 

\section{Related Work}\label{sec_relwork}
As described in Section \ref{sec_intro}, Bogdiukiewicz et al. proposed a method to develop a policing function for autonomous systems \cite{policing}. The policing function checks output values of intelligent function such as a machine learning model at runtime. The policing function is useful to prevent failures as well as the input filter. However, the policing function works after the intelligent function executes. It means that the policing function detects and controls the possible failure later than the input filter. Therefore, the proposed method for creating the input filter is more useful in develepment of systems which require quick handling of failures.

In $formal\_verification$ shown in Section \ref{sec_fv}, a rule-form formula composed of conditions and conclusions is extracted from the DTEM. 
Extracting a specification from a program in the form of rules has been studied \cite{cfg_analysis1}\cite{cfg_analysis2}\cite{cfg_analysis3}\cite{cfg_analysis4}\cite{cfg_analysis5}\cite{cfg_analysis6}\cite{symb_exec1}. 
In these studies, the control path of the program is extracted by static analysis or symbolic execution, and the rule is created on the basis of the branch condition that constitutes the control path. The method proposed in this paper extracts the decision tree path and creates a rule-form formula based on the attribute test associated with the decision node that composes the path. It can thus be said that it adopts the same approach.

As for the verification of the rule-form formula, methods have been established to verify properties such as consistency, redundancy, and completeness \cite{veri_rule}. Furthermore, a method for extracting ``minimal unsatisfiable subsets," which are useful for causal analysis when the rule set does not satisfy consistency, has also been proposed \cite{mus1}\cite{mus2}\cite{mus3}. Validation of rule programs used in a business rule management system has also been studied \cite{brms}. As in the case of the method proposed in this paper, the verification of the the rule-form formula is translated to checking the satisfiability problem.

Moreover, as described in Section \ref{sec_intro}, an approach of encoding a DNN model in a formula and verifying it by determining the satisfiability of the formula has been proposed in recent years \cite{stanford_lp1}\cite{stanford_lp2}\cite{stanford_lp3}\cite{planet}\cite{oxford_smt}\cite{unified}. It can be said that the method for verifying the DTEM proposed in this paper also takes this approach. However, a study describing a specific method for verifying a DTEM has not been reported. One of the contributions of this paper is formally specifying a verification method for a DTEM and demonstrating its feasibility through a case study.

No similar studies on extracting violation ranges as described in Section \ref{sec_extract} and dividing violation ranges as described in Section \ref{sec_divide} have been reported. One of the reasons for that situation is that the problems solved by these methods are unique to software developed by machine learning. In regard to conventional software, namely, algorithmic programs, when a counterexample is detected by verification, the fault that caused the counterexample is analyzed, and the fault is removed by correcting the algorithm. Verification and correction are then repeated until no counterexample is detected. On the other hand, in the case of a machine learning model, a possible method for handling such faults is retraining or additional training using the detected counterexample (and data similar to it). Although this approach may eliminate the fault, retraining and additional training may affect the entire model, and another new fault may be inserted. That is, ``regression" occurs. Although this regression also occurs in the case of algorithmic programs, in that case, regression occurs due to a developer's mistake; therefore, such mistakes must be careful corrected to avoid regression. On the contrary, in the case of a machine-learning model, it is difficult for the developer to control retraining and additional training so that regression does not occur. In other words, it can be said that it is inherently difficult to create a complete machine-learning model that always returns the expected output value. Accordingly, when a machine-learning model is implemented in a system, the proposed method is used to implement the input filter. That is, the proposed method solves a specific problem that occurs with the use of machine-learning models.

\section{Conclusion}\label{sec_conclusion}
When a DTEM is implemented in a system, the input filter can be effectively used to prevent system failures. As a means of creating the filtering condition for the input filter, a method for extracting the violation range of the DTEM, whose input values are multi-dimensional vectors whose elements are numeric variables, was proposed. The proposed method consists of procedures for formally verifying the DTEM, extracting the violation range, and narrowing the extracted violation range. The violation range extracted by the proposed method includes all violating input values. The proposed method is therefore useful to create the filtering condition. After showing the algorithms for these procedures, the results of a case study using a dataset on the house prices were presented. 
On the basis of the results of this case study, the feasibility of the proposed method is demonstrated. Through the case study, the scalability of the proposed method is also evaluated. 
The number of decision trees constituting the DTEM, the maximum depth of the decision trees, and the dimension of the input value are factors that increase the execution time of the proposed method. Specifically, it was concluded that the proposed method is practical if the number of decision trees is less than around 200 and the maximum depth of the decision trees is less than 5 at least.

Future issues include improving the scalability of the proposed method. Since the form of the formula $F_{M}$ created by the proposed method is constant, the method's scalability may be improved by implementing heuristics specialized for that form in the SMT solver. Moreover, the procedure for finding appropriate values of parameters $r_{a}$, $r_{b}$, and $r_{c}$ shown in Section \ref{sec_discuss} can be incorporated into the proposed method. Furthermore, by adding case studies using other datasets, the limitation of the proposed method can be evaluated in more detail.


\begin{thebibliography}{99}

\bibitem{dl_image} A. Krizhevsky, I. Sutskever, and G. Hinton. Imagenet Classication with Deep Convolutional Neural Networks. Advances in Neural Information Processing Systems, pages 1097-1105, 2012.

\bibitem{dl_speech} G. Hinton, L. Deng, D. Yu, G. Dahl, A. Mohamed, N. Jaitly, A. Senior, V. Vanhoucke, P. Nguyen, T. Sainath, and B. Kingsbury. Deep Neural Networks for Acoustic Modeling in Speech Recognition: The Shared Views of Four Research Groups. IEEE Signal Processing Magazine, 29(6):82-97, 2012.

\bibitem{rf} L. Breiman: Random forests. Maching Learning, 45(1), 5-32, (2001).

\bibitem{gbdt} J.H. Friedman: Greedy function approximation: a gradient boosting machine. Annals of statistics, pp.1189-1232 (2001).

\bibitem{gbdt_ex1} P. Li: Robust logitboost and adaptive base class (abc) logitboost. arXiv preprint arXiv:1203.3491 (2012).

\bibitem{gbdt_ex2} M. Richardson, E. Dominowska, and R. Ragno: Predicting clicks: estimating the click-through rate for new ads. In Proceedings of the 16th international conference on World Wide Web, pp.521-530. ACM (2007).

\bibitem{gbdt_ex3} C.J.C. Burges: From ranknet to lambdarank to lambdamart: An overview, Microsoft Research Technical Report MSR-TR-2010-82 (2010).

\bibitem{rf_ex1} R.A.V. Rossel, and T. Behrens: Using data mining to model and interpret soil diffuse reflectance spectra, Geoderma 2010, 158, pp.46-54 (2010).

\bibitem{rf_ex2} P.T. Sorenson, C. Small, M.C. Tappert, S.A. Quideau, B. Drozdowski, A. Underwood, and A. Janz: Monitoring organic carbon, total nitrogen, and pH for reclaimed soils using field reflectance spectroscopy. Can. J. Soil Sci. 2017, 97, 241-248 (2017).

\bibitem{rf_ex3} A.M. Prasad, L.R. Iverson, and A. Liaw: Newer classification and regression tree techniques: Bagging and random forests for ecological prediction. Ecosystems 2006, 9, pp.181-199 (2006).

\bibitem{rf_ex4} H. Ishwaran: Variable importance in binary regression trees and forests. Electron. J. Stat. 2007, 1, pp.519-537 (2007).

\bibitem{policing} C. Bogdiukiewicz, M. Butler, T.S. Hoang, M. Paxton, J.H. Snook, X. Waldron, and T. Wilkinson: Formal Development of Policing Functions for Intelligent Systems, In 2017 IEEE 28th International Symposium on Software Reliability Engineering (2017).

\bibitem{xgboost} T. Chen, and C. Guestrin: Xgboost: A scalable tree boosting system. In Proceedings of the 22Nd ACM SIGKDD International Conference on Knowledge Discovery and Data Mining, pp.785-794. ACM, (2016).

\bibitem{oxford_smt} X. Huang, M. Kwiatkowska, S. Wang, and M. Wu.: Safety Verification of Deep Neural Networks, Computer Aided Verification 2017, Lecture Notes in Computer Science, vol.10426, pp.3-29 (2017).

\bibitem{stanford_lp1} G. Katz, C. Barrett, D.L. Dill, K. Julian, and M.J. Kochenderfer: Reluplex: An Efficient SMT Solver for Verifying Deep Neural Networks, Computer Aided Verification 2017, pp.97-117 (2017).

\bibitem{stanford_lp2} G. Katz, C. Barrett, D.L. Dill, K. Julian, and M.J. Kochenderfer: Towards Proving the Adversarial Robustness of Deep Neural Networks, arXiv:1709.02802 (2017).

\bibitem{stanford_lp3} D. Gopinath, G. Katz, C. Pasareanu, and C. Barrett: DeepSafe: A Data-driven Approach for Checking Adversarial Robustness in Neural Networks, arXiv:1710.00486 (2017).

\bibitem{planet} R. Ehlers:  Formal verification of piece-wise linear feed-forward neural networks. Automated Technology for Verification and Analysis 2017 (2017).

\bibitem{unified} R. Bunel, I. Turkaslan, P.H.S. Torr, P. Kohli, and M.P. Kumar: A Unified View of Piecewise Linear Neural Network Verification, arXiv:1711.00455 (2018)

\bibitem{decisiontree1} P.E. Utgoff: Incremental induction of decision trees, Machine Learning 4: 161.(1989). \url{https://doi.org/10.1023/A:1022699900025} 

\bibitem{decisiontree2} A.M. Bhavitha S, and S. Madhuri: A Classification Method using Decision Tree for Uncertain Data, International Journal of Computer Trends and Technology (IJCTT),V3(1), pp.102-107 (2012).


\bibitem{cfg_analysis1} T. Hatano, T. Ishio, J. Okada, Y. Sakata, and K. Inoue: Extraction of Conditional Statements for Understanding Business Rules, Proceedings of 6th International Workshop on Empirical Software Engineering in Practice (2014).

\bibitem{cfg_analysis2} V. Cosentino, J. Cabot, P. Albert, P. Bauquel, and J. Perronnet: Extracting business rules from COBOL: A model-based framework, 20th Working Conference on Reverse Engineering (WCRE) pp.409-416 (2013).

\bibitem{cfg_analysis3} H.M. Sneed: Extracting Business Logic from Existing COBOL Programs as a Basis for Redevelopment，Proceedings of the 9th International Workshop on Program Comprehension pp.167-175 (2001).

\bibitem{cfg_analysis4} X. Wang, J. Sun, and X. Yang: Business rules extraction from large legacy systems, in Proc. CSMR, 2004, pp.249-253 (2004).

\bibitem{cfg_analysis5} H. Huang and W. Tsai: Business rule extraction from legacy code, in Proc. COMPSAC, 1996, pp.162-167 (1996).

\bibitem{cfg_analysis6} V. Cosentino, J. Cabot, P. Albert, P. Bauquel, and J. Perronnet: A Model Driven Reverse Engineering Framework for Extracting Business Rules out of a Java Application, in Proc. RuleML, 2012, pp.17-31 (2012).

\bibitem{symb_exec1} J. Pichler: Specification extraction by symbolic execution, in Proc. of 20th Working Conference on Reverse Engineering, pp.462-466 (2013).

\bibitem{symb_exec2} R. Baldoni, E. Coppa, D. C. D'elia, C. Demetrescu, and I. Finocchi: A Survey of Symbolic Execution Techniques, ACM Computing Surveys (CSUR), v.51 n.3, pp.1-39 (2018).



\bibitem{veri_rule} A. Ligeza, G.J. Nalepa: Rules verification and validation, Handbook of Research on Emerging Rule-Based Languages and Technologies: Open Solutions and Approaches, pp.273-301 (2009).

\bibitem{mus1} J. Bailey and P.J. Stuckey: Discovery of minimal unsatisable subsets of constraints using hitting set dualization, In Practical Aspects of Declarative Languages, 7th International Symposium, PADL 2005, Lecture Notes in Computer Science, vol.3350, pp.174-186 (2005).

\bibitem{mus2} M.H. Liffiton and A. Malik: Enumerating infeasibility: Finding multiple MUSes quickly, In Integration of AI and OR Techniques in Constraint Programming for Combinatorial Optimization Problems, CPAIOR 2013, Lecture Notes in Computer Science, vol.7874, pp.160-175 (2013).

\bibitem{mus3} M.H. LiffitonEmail and K.A. Sakallah: Algorithms for computing minimal unsatisable subsets of constraints, Journal of Automated Reasoning, Volume 40, Issue 1, pp.1-33 (2008). 

\bibitem{mus4} T.S. Hoang, S. Itoh, K. Oyama, K. Miyazaki, H. Kuruma, and N. Sato: Consistency Verification of Specification Rules, Formal Methods and Software Engineering (ICFEM) 2015, Lecture Notes in Computer Science, vol.9407 pp.50-66 (2015).





\bibitem{brms} B. Berstel and M. Leconte: Using constraints to verify properties of rule programs, In Third International Conference on Software Testing, Verifcation and Validation, ICST 2010, pp.349-354 (2010).





\bibitem{z3} L. de Moura and N. Bjorner:Z3: An Efficient SMT Solver, Tools and Algorithms for the Construction and Analysis of Systems. TACAS 2008. Lecture Notes in Computer Science, vol 4963. pp.337-340 (2008).




\end{thebibliography}
\end{document}